\documentstyle[preprint,aps]{revtex}
\draft
\tighten
\begin{document}
\title{
Effect of screening of the Coulomb interaction on the conductivity
in the quantum Hall regime
}
\author{I.\ L.\ Aleiner and B.\ I.\ Shklovskii}
\address{
Theoretical Physics Institute,
School of Physics and Astronomy, University of Minnesota,\\
116 Church St. SE, Minneapolis, MN 55455}
\maketitle

\begin{abstract}
We study variable range hopping in the quantum Hall effect regime
in the presence of a metallic gate parallel to the plane of a
two-dimensional electron gas. Screening of the Coulomb interaction
by the gate
causes the partial ``filling'' of the Coulomb gap in the
density of localized states. At low enough temperatures
this leads to a substantial enhancement and a new temperature behavior
of the hopping conductivity. As a result, the diagonal conductivity peaks
become much wider.
The power law dependence
of the width of the peaks on the temperature changes:
the corresponding exponent turns out to be twice  as small
as that for  gateless structures. The width dependences on the current in
non-ohmic regime and on the frequency for the absorption of the
electromagnetic waves experience a similar modification.
The experimental observation of the
crossovers predicted may demonstrate the important role
of the Coulomb interaction in the integer quantum Hall regime.
\end{abstract}
\narrowtext

\section{Introduction}
\label{Intro}
The integer quantum Hall effect (IQHE)
in a disordered two-dimensional electron gas
manifests itself more clearly as the temperature $T$ is lowered.
The steps
connecting adjacent plateaus in the dependence of the Hall conductance
$\sigma_{xy}$ on the filling factor $\nu$ narrow with decreasing $T$ and so do
the peaks in the longitudinal conductance $\sigma_{xx}$. In a number of
experiments
\cite{wei88,wei92,koch91,koch92,dolgopolov91,iorio92} a
remarkable result has been obtained: the width $\Delta\nu$ of
the peaks shrinks as $T\to 0$ according to a power law
$\Delta\nu\propto T^\kappa$ in measurements having been
performed down to temperatures as low as a few tens of millikelvins.
The exponent $\kappa\simeq 0.4$ was
found in Refs.~\cite{wei88,wei92} to be universal: neither the
Landau level index nor the electron mobility is relevant at low
temperatures. However, some deviations from $\kappa\simeq 0.4$ were reported
in Refs.\cite{koch91,koch92}.

A new explanation of the scaling dependence
$\Delta\nu\propto T^\kappa$ has been recently suggested  by Polyakov
and Shklovskii (PS)\cite{Polyakov}. (Similar arguments with respect to
insulator-superconductor transition were put forward earlier by
Fisher\cite{Fisher1}.)
They noticed the fact that
in the low-temperature limit  the
only possible mechanism of transport away from the peaks where the states
are localized is
variable-range hopping (VRH). In this regime, due to the existence of the
Coulomb gap, the temperature dependence of $\sigma_{xx}$ should have a form
\cite{efros75,shklovskii84,efros85+}
\begin{equation}
\sigma_{xx}=\sigma_0e^{-(T_0/T)^{1/2}}, \label{hopping}
\end{equation}
where
\begin{equation}
T_0(\nu)=C{1\over k_B}{e^2\over \varepsilon\xi(\nu)}, \label{T_0}
\end{equation}
$\xi(\nu)$ is the localization radius
of the states on the Fermi level for a given $\nu$, $\varepsilon$ is the
dielectric constant, $k_B$ is the Boltzmann constant, and
$C\simeq 6.2$ in
two dimensions \cite{nguyen84}.

Numerical simulations
\cite{aoki85,chalker88,huckestein90,ando92,huo92}
show that the localization length
$\xi(\nu)$ diverges as $\nu$ approaches a half-integer $\nu_0$:
\begin{equation}
 \xi(\nu)=\xi_0|\nu-\nu_0|^{-\gamma},
\label{localization}
\end{equation}
where $\xi_0$ is of the order of the magnetic length. For the case of
spinless electron which corresponds to the spin split Landau level
the value of the exponent $\gamma$  was shown to be given by
\cite{aoki85,chalker88,huckestein90,ando92,huo92}
\begin{equation}
 \gamma = \gamma_0\simeq 2.3 .
\label{gamma0}
\end{equation}
Recently, the same value of
$\gamma$ has been directly measured by studying how $\Delta\nu$ scales with the
sample size in the low-$T$ limit \cite{koch92}.

According to Eqs.~(\ref{T_0}), (\ref{localization}),
the value of $T_0$ tends to
zero as $\nu\to \nu_0$. Hence, at a given temperature, there should exist a
characteristic value of $\nu$ at which  $T_0(\nu)\sim T$
and the exponential factor in
Eq.~(\ref{hopping}) becomes of the order of unity. PS assumed
that the width of the peak is determined by the difference between
this value and $\nu_0$.
Making use of  Eq.~(\ref{localization})
immediately yields a
power-law dependence of $\Delta\nu$ on $T$:
\begin{equation}
\Delta\nu=\left( {T\over T_1}\right) ^\kappa
\label{nu_t}
\end{equation}
with $\kappa=1/\gamma$ and
\begin{equation}
T_1=A{1\over k_B}{e^2\over \varepsilon\xi_0},
\label{T_1}
\end{equation}
where $A$ is a numerical coefficient.
For $\gamma\simeq 2.3$ one obtains $\kappa\simeq 0.4$
which is in a good agreement with the experimental data.

The approach by PS can be related to
the conventional theory of the width of the conductivity peaks which was
proposed by Aoki and Ando \cite{aoki85} and by Pruisken\cite{Pruisken}.
This theory is based upon the concept of a phase-coherence length
$L_{\phi}$ which sets a limit for the localization. If $L_{\phi}$
is less than the localization
length $\xi$ of the state at the Fermi level, the
state is considered to be delocalized and it contributes
to the conductivity.
Then, the width
of the conductivity peak should be determined from condition
$L_{\phi} \simeq \xi$. Using this concept, one arrives at
Eqs.~(\ref{nu_t}),~(\ref{T_1}) if the  phase-coherence length is
given by
\begin{equation}
 L_{\phi} \simeq \frac{e^2}{\varepsilon k_BT}.
 \label{LFisher}
\end{equation}
Earlier for the case of the Coulomb interaction the same
expression was suggested by Fisher, Girvin and Grinstein \cite{Fisher}
in the framework of dynamic scaling theory. Additional arguments
for Eq.~({\ref{LFisher}) have been given recently by Lee, Wang and
Kivelson~\cite{Kivelson}.

The discussion above shows an important role
of the long range Coulomb correlations even in low mobility samples.
These correlations create the Coulomb gap at the density of states
suppressing the hopping conductivity between the peaks and
causing the substantial modification of the conductivity
peaks shapes.

The large spatial scale of the correlations allows one
to examine the role of the Coulomb interaction experimentally by making
use of the intentionally introduced screening.

In this paper we concentrate on  one of the possibilities of
such a screening, namely on  study of gated structures,
i.e.  structures with a metallic layer being placed parallel to
the plane of the 2D electron gas at a distance $d$. First measurements
of the IQHE on the GaAs gated structures have been already reported
\cite{Jiang,Dahm,Hughes,Friedland}.

In Sect.~\ref{sec:density} we start from the discussion of how the screening
of the interaction affects
the shape of the density of states in the classical
limit (i.e. for the localization length of the states smaller
than the mean distance between them). The main effect of the
screening is the appearance of a non-vanishing value
of the density of states at the Fermi level. We use the
results of Mogilyanskii and Raikh \cite{Raikh} who studied
this case quantitatively. We will argue that the results
remain valid even in the IQHE  regime where the localization length
$\xi$ can be much greater than the distance between the states
which is of the order of $\lambda$.

In Sect.~\ref{sec:temperature} we use these results
to derive the temperature dependence
of the conductivity in the VRH regime. This dependence appears to
have a crossover from the Eq.~(\ref{hopping}) to  Mott's
law
\begin{equation}
\sigma_{xx}=\sigma_Me^{-(T_M/T)^{1/3}}, \label{Mott}
\end{equation}
where $T_M$ is inversely proportional to the density of states at
the Fermi level. The density of states, however, is strongly diminished
by the Coulomb interaction at distances smaller than $d$. Making
use of Eq.~(\ref{Mott}), we extend PS approach to show
that the temperature dependence of the
width of the conductivity peaks
has a crossover with decreasing temperature
from Eqs.~(\ref{nu_t}), (\ref{T_1}) to
\begin{equation}
\Delta\nu=\left( {T\over T_2}\right)^{\kappa},
\label{nu_t_mott}
\end{equation}
where
\begin{equation}
\kappa=1/2\gamma, \quad T_2=B{1\over k_B}{e^2d\over \varepsilon\xi_0^2}.
\label{T_2}
\end{equation}
The numerical coefficient $B$ in the Eq.(\ref{T_2}) is
approximately equal to $140$. Crossover happens at temperature
\begin{equation}
 T_c \simeq 0.25\frac{e^2}{k_B\varepsilon d}
 \label{T_c}
\end{equation}
if $d \gg \xi_0$. In the opposite case $d \ll \xi_0$ the width
of the peak is given by Eq.~(\ref{nu_t_mott}) in the whole range
$\Delta\nu < 1/2$. The function $\Delta\nu(T)$ at different values of
$d$ is shown schematically in Fig.~1. Observation of such a crossover
in scaling dependences would provide evidence for the role
of the Coulomb interaction.

PS applied their approach also to get  dependences
of the peak width  upon
the frequency $\omega$ and current $J$.
These dependences were shown to have the power
law form $\Delta\nu\propto J^{\mu}$ and $\Delta\nu\propto \omega^{\eta}$.
The same phenomena for the gated structures
are addressed in Sections \ref{sec:current} and \ref{sec:frequency}.
We show that the exponent $\mu$
experiences the crossover between $\mu=1/2\gamma$ and $\mu=1/3\gamma$
with  decreasing current.
With decreasing frequency, crossover happens between $\eta=1/\gamma$ and
$\eta=1/2\gamma$.

In Sec.~\ref{sec:experiment}
we discuss the existing experimental data for GaAs gated
structures and Si MOSFETs.

\section{Effect of the gate on the density of states in the Coulomb gap}
\label{sec:density}

The variable range hopping is  known to depend strongly
on the shape of the density states at the energies  close to the
Fermi level. If the Coulomb interaction  at  large distances
is not screened by the gate, the density of states
$g_C$ is of the form\cite{shklovskii84}
\begin{equation}
 g_C(E)=\frac{2}{\pi}\frac{\varepsilon^2}{e^4}|E|,
 \label{ds}
\end{equation}
with the energy $E$ being counted from the Fermi level.

It is the long range nature of the Coulomb interaction
that makes the density of states vanish at the Fermi level.
Therefore, it is natural to expect that the screening of the
interaction at
large distances leads to the
finite value of the density of states  at the Fermi
level\cite{shklovskii84,efros85+,raikh2}.
Indeed, the dependence (\ref{ds}) is
determined by the interaction at distances of the order of
$r_E=\frac{e^2}{\varepsilon |E|}$. If the potential is
screened at the distances greater than $r_s$ (i.e. the energy of the
interaction decays faster than $1/r^2$ at $r > r_s$), the
interaction between two electrons of energies
$|E| < E_s\equiv\frac{e^2}{\varepsilon r_s}$ can be neglected.
But this interaction is the only reason for the density
of states to be suppressed at the Fermi level.
Thus, the energy dependence of the density of states saturates
at  the energy of the order of $E_s$, so that $g(0) \simeq g_C(E_s)$
with $g_C(E)$ given by Eq.~(\ref{ds}).

In the case of gated structures, the
Coulomb potential is screened by the gate located
at the distance $d$ from the plane of 2D electron gas so
that the energy of the electron-electron interaction has
a form
\begin{equation}
 U(r) = \frac{e^2}{\varepsilon}\left(
 \frac{1}{r} - \frac{1}{\sqrt{r^2 + 4d^2}}
 \right).
 \label{dipole}
\end{equation}
In this case distance $d$ plays  the role of the screening radius
$r_s$, $E_s \simeq \frac{e^2}{\varepsilon d}$ and the density of
states at the Fermi level is given by
\begin{equation}
 g(0) = \alpha\frac{\varepsilon}{e^2 d},
 \label{g0}
\end{equation}
where $\alpha$ is a numerical coefficient which is
to be determined from
more rigorous calculations. Such calculations have been done
numerically
by Mogilyanskii and Raikh \cite{Raikh}. Using self-consistent
approach developed in Ref.\cite{raikh3}, they obtained the anomalously
small value of the coefficient $\alpha \approx 0.1$. Furthermore,
they suggested the following interpolation  formula fitting the results of
numerical
simulations of the density of states for the interaction potential
(\ref{dipole}):
\begin{equation}
 g(E)=\frac{\varepsilon^2}{e^4}\left(\frac{2}{\pi}|E| +
 \alpha\frac{e^2}{\varepsilon d}\right).
 \label{screen}
\end{equation}

Eqs.~(\ref{ds}), (\ref{screen}) have been derived for the
classical limit with localization length of the states
$\xi$ being much smaller than the mean distance between them.
It is not the case for the filling factor of the Landau level close to
a half integer  when according  to Eq.~(\ref{localization})
the localization length at the Fermi level diverges whereas
the mean distance between electrons is of the order of the magnetic length
$\lambda$. It is easy to realize that
the energy of the interaction of two
localized states is  given by the Coulomb law only
if the distance between them exceeds the localization length $\xi$.
At smaller distances, the interaction energy saturates at
the value
of the order of $E_{\xi}=\frac{e^2}{\varepsilon\xi}$.
Such a form of the interaction
energy would lead to a step-like increase of the density of states
at energies larger than $E_{\xi}$. In other words,
a ``hard'' gap of width of the order of
$E_{\xi}$ would exist at such energies rather than
the ``soft'' gap  given by Eq.~(\ref{ds}).

However, we believe that, if  dielectric constant $\varepsilon$
does not grow with spatial scale\cite{Polyakov},
the ``soft'' Coulomb gap survives
at these energies, too.
We would like to mention two papers which support this point of view.
Yang and MacDonald\cite{macdonald} performed self consistent
Hartree-Fock calculations of the system of interacting electrons in a strong
magnetic field in the presence of  disorder.  They demonstrated the existence
of the Coulomb gap even for the Fermi level close to the center
of first Landau level. Recently, some theoretical arguments for the Coulomb
gap in under these conditions were provided in Ref.~\cite{Aleiner}
where we have
argued that
at the energies $|E| > E_{\xi}$ the one-electron wave functions
of the Hartree-Fock approximation shrink because of
the electron-electron interaction so that their localization length
$\zeta (E)$ is determined by $\zeta (E)\geq\frac{e^2}{\varepsilon |E|}$
rather than by the one-electron value $\xi$. States closer to the Fermi
level ($|E| < E_{\xi}$) still have  one-electron localization length
$\xi$.
Such a picture self-consistently leads to
the existence of the Coulomb gap.
Indeed, if Coulomb gap exists, the average distance between states in the
band of  width $E$ around
the Fermi level is of the order of $\frac{e^2}{\varepsilon |E|}$ which
is not smaller than the localization length of the states  $\zeta (E)$. Thus,
one  can use the Coulomb interaction energy and arrive back
to Eq. (\ref{ds}), which
makes the theory self-consistent.
This picture allows us to make use of the Eqs.~(\ref{ds}),
(\ref{screen}) even though the one electron $\xi$
is large and to treat the Coulomb interaction effects classically.

Below we repeat our arguments\cite{Aleiner} for the phenomenon
of the additional localization of Hartree-Fock functions.
We showed that the  system in which
all the one electron states have the one electron  localization length
$\xi$
is unstable with respect to the
following transformation. Let us construct wave-packets of the
size $L\equiv\frac{e^2}{\varepsilon|E|}$ from the one-electron states.
The necessary work is of the order of $(g_0L^2)^{-1}$ per packet, where
$g_0$ is the density of states for non-interacting electrons.
On the other hand, due to the electron-electron interaction,
the system will gain from crystal-like Coulomb
correlations of the wave packets an energy
of the order of $\frac{e^2}{\varepsilon{L}}$ per packet.
This gain is larger than
$(g_0L^2)^{-1}$ if $|E|<\frac{g_0 e^4}{\varepsilon^2}$.
It means that the Eq.~(\ref{ds}) is valid in the
whole range $|E|<\frac{g_0 e^4}{\varepsilon^2}$  and that
the localization length for  $|{E}|>\frac{e^2}{\varepsilon\xi}$
is about $\frac{e^2}{\varepsilon|E|}$.

It is important to emphasize once more
that these arguments essentially rely on
assumption\cite{Polyakov} that in a strong magnetic field  dielectric
constant $\varepsilon$
does not diverge with $\xi$. This situation differs from the case
of zero magnetic field in which $\varepsilon$ does diverge\cite{larkin}.

\section{Temperature dependence of the width of the peaks of
the diagonal conductivity}
\label{sec:temperature}

The modification of the density of states
in the gated structure results in a qualitative change of
the temperature dependence in  VRH. In the limit of
low temperatures the activation energy  for a typical hop
contributing to the conductivity   is much
less than $\alpha\frac{e^2}{\varepsilon d}$ and the length of such a hop
is much larger than $d$. It means that
the Coulomb interaction is not relevant  in this regime for the
hopping itself, though it affects the value of the density of states
(see Sec.~\ref{sec:density}). Therefore, in
this temperature range, the hopping conductivity obeys
Mott's formula~(\ref{Mott}) with
\begin{equation}
 T_M=\frac{\beta}{k_Bg(0)\xi^2},
\label{T_M}
\end{equation}
where $g(0)$ is given by Eq.~(\ref{g0})
and the numerical coefficient $\beta\simeq 14$\cite{shklovskii84}.

In the whole range of temperature  VRH is described by
\begin{equation}
\sigma_{xx}=\sigma_0
\exp\left(-\left\{\frac{T_0f(T/T_*)}{T}\right\}^{1/2}
\right).
\label{wholerange}
\end{equation}
Here, $T_0$ is given by Eq.~(\ref{T_0}),
the characteristic temperature $T_*$ is defined as
\begin{equation}
  T_*\equiv \frac{e^2\xi}{k_B\varepsilon d^2},
 \label{critical T}
\end{equation}
and $f(x)$ is a dimensionless function. This
function can be obtained
in the framework of the concept of invariance of the number
of connections per site in the
percolation problem\cite{shklovskii84}. The
result is shown
in Fig. 2a.
The asymptotic behavior of $f(x)$ that can be obtained immediately
from Eqs.~(\ref{hopping}), (\ref{Mott}), (\ref{wholerange}) and has
a form:
\begin{equation}
 f(x) = \left\{
    \begin{array}{ll}
        1, & \mbox{if $x \gg 1$}, \\
        \displaystyle{\frac{\beta^{2/3}x^{1/3}}{C\alpha^{2/3}}}
        \simeq\sqrt[3]{78x}, & \mbox{if $x \ll 1$},
    \end{array}
 \right.
 \label{f}
\end{equation}
the coefficients $\alpha$ and $C$ being defined in Eqs.~(\ref{g0}) and
(\ref{T_0}) respectively.
Crossover between the asymptotes (\ref{f}) which are also shown in
Fig.~2a
happens near $x \simeq 0.013$, i.e. around
$T = \tilde{T} \simeq 0.013 T_*$.
By studying the crossover experimentally one can determine
the localization length $\xi$ as a function of $\nu$.

Let us now turn to derivation of the
temperature dependence of the width of the
diagonal conductivity peaks. In the spirit of PS approach, this dependence
can be obtained by equating the term in the exponent in
Eq.~(\ref{Mott}) or in Eq.~(\ref{wholerange}) to the unity.
Making use of Eq.~(\ref{T_M}), we obtain for  the lowest
temperatures
\begin{equation}
 \xi^2(\Delta\nu)=\frac{140e^2d}{\varepsilon k_BT},
\label{Boris}
\end{equation}
which with help of Eq.~(\ref{localization}) yields Eq.~(\ref{nu_t_mott}).
In the whole range of temperatures the temperature dependence
of the width of the peak can be written as:
\begin{equation}
\Delta\nu=\left\{\frac{\xi_0}{d} F(T/T_d)\right\}^{1/\gamma}
\label{nu_t1}
\end{equation}
with $T_d$ given by
\begin{equation}
T_d = \frac{e^2}{k_B\varepsilon d}.
\label{T_d}
\end{equation}
Dimensionless function $F(x)$ obtained as the solution of the equation
\begin{equation}
 C f(xF(x))= \frac{x}{F(x)}
\label{g}
\end{equation}
is depicted in Fig.~2b.
Making use of Eq.~(\ref{f}) allows us to derive the asymptotic
behavior of the function $F(x)$
\begin{equation}
   F(x) = \left\{
    \begin{array}{ll}
     x/C \simeq x/6.2, \quad & \mbox{if $x \gg 1$}; \\
     \sqrt{\alpha x/\beta} \simeq \sqrt{x/140},\quad & \mbox{if $x \ll 1$},
    \end{array}
 \right.
 \label{g_as}
\end{equation}
which corresponds to Eqs.~(\ref{nu_t}), (\ref{nu_t_mott}). The crossover
between two asymtotics happens at $x \simeq 0.25$.

Expression (\ref{nu_t1}) is meaningful only if $\Delta\nu \ll 1$ and scaling
formula (\ref{localization}) is valid. Thus, the crossover is observable
only if $d \gtrsim 0.05 \xi_0$, what is satisfied in most of
realistic situations. Otherwise, the whole dependence
is described by Eq.~(\ref{nu_t_mott}).

The crossover can be interpreted in the framework of the dynamical
scaling approach of Ref.\cite{Fisher}.
According to this approach, the characteristic time $\hbar/T$ scales as
$\xi^z$. At low temperatures the relevant
spatial scales are large and all  Coulomb interaction is screened
by the gate. Obtained for this case
Eq.~(\ref{Boris}) corresponds to $z=2$
in agreement with the result
of Ref.\cite{Fisher,Fisher2}. With temperature
increasing the spatial scale decreases and
becomes less than the distance to the gate $d$.
At these scales  long-range nature of the
Coulomb interaction becomes relevant
and results in the dynamical exponent $z=1$.

\section{Effect of gate on current induced broadening of the conductivity
peaks}
\label{sec:current}

It was found in Refs.\cite{koch92,simmons91}
that the width $\Delta\nu$ of the $\sigma_{xx}$ peaks grows with increasing
current $J$, i.e. with the increase of the Hall electric field ${\cal E}_H$.
PS showed
\cite{Polyakov} that the dependence $\Delta\nu({\cal E}_H)$
can be understood in terms
of the theory of hopping in a strong electric field
\cite{shklovskii90}.
This theory is based on the fact that there
exists a quasi-Fermi level inclined by the electric field $\cal E$.
Zero-temperature hopping with phonon emission becomes then possible and, even
though there are no absorption processes,
the local Fermi distribution with an
effective temperature $\sim e\cal E\xi$ is formed
\cite{shklovskii90}. On this account, the exponent of
the current-voltage characteristics at $T=0$ may be estimated from
that of the
ohmic conductivity by replacing  $T\to e{\cal E}\xi$.
If the ohmic transport obeys the law (\ref{Mott}), the zero-temperature
conductivity should behave with increasing electric field as
\begin{equation}
\sigma_{xx} \propto e^{-({\cal E}_M/{\cal E}_H)^{1/3}}, \quad {{\cal
E}_M} \simeq {k_BT_M\over e\xi}.
\end{equation}
Similarly to the case of ohmic conductivity the width of the $\sigma_{xx}$
peak is found from the equation ${\cal E}_M(\xi)\sim {\cal E}_H$.
Solving this equation for $\xi$ we get
$\xi \sim \left(\frac{\beta}{g(0)e{\cal E}_H}\right)^{1/3}$
($\beta$ is a numerical coefficient from Eq.~(\ref{T_M})), which
with help of Eqs.~(\ref{screen}) and (\ref{localization}) yields
\begin{equation}
\Delta\nu={\left( {{\cal E}_H\over {\cal E}_1}\right)}^{\mu}
={\left( {J\over
J_1}\right)}^{\mu},
\label{nu_E}
\end{equation}
where
\begin{equation}
\mu=\frac{1}{3\gamma},
{\cal E}_1=B\frac{\beta e d}{\alpha\varepsilon\xi_0^3},
J_1=\sigma_{xy}(\nu_0){\cal E}_1,
J = \sigma_{xy}(\nu_0){\cal E}_H,
\label{E1}
\end{equation}
$B$ being a numerical coefficient of the order of unity. These results
should be compared with the corresponding results for the gateless
structures\cite{Polyakov}:
\begin{equation}
\mu=\frac{1}{2\gamma},
\quad {\cal E}_1= B_1\frac{Ce}{\varepsilon\xi_0^2},
\quad J_1\sim \sigma_{xy}(\nu_0){{\cal E}_H}_1,
\label{E2}
\end{equation}
$B_1$ being a coefficient of the order of unity.
We see that the screening of the Coulomb interaction by gate
substantially affects the character of the current broadening and
changes the exponent $\mu$.

Similar to the case of the temperature broadening, crossover between
two exponents $\mu=1/3\gamma$ and $\mu=1/2\gamma$ in Eq.~(\ref{nu_E})
takes place with increasing current.
Comparing Eqs.~(\ref{E1}) and (\ref{E2}), we obtain that this crossover
happens at
\begin{equation}
 J_c \simeq 0.01 \sigma_{xy}(\nu_0) \frac{e}{\varepsilon d^2}.
\end{equation}

\section{Effect of gate on broadening of the {\it ac-} conductivity
  peaks with frequency}
\label{sec:frequency}

In this section  we deal with  the ohmic zero-temperature conductivity in the
quantum Hall regime at a finite frequency $\omega$.
For the gateless structures broadening of narrow
$\sigma_{xx}$-peaks has been observed  at a few tens millikelvins as the
frequency changed in the range $\sim 0.2-15\: \mbox{GHz}$ \cite{omega}.
This phenomenon has been also interpreted in
terms of hopping conductivity\cite{Polyakov}. In contrast to the {\it
dc} case, the hopping conductivity at a  frequency smaller
than the mean energy level spacing $\Delta_{\xi}$ in a
square of the size $\xi$, is determined by
sparsely distributed pairs of the localized states, the typical separation
between two sites of a pair being much shorter than that between pairs.
The main contribution
to the conductivity is provided by resonant phononless
transitions of the electrons from one site of a pair to another if
$\hbar\omega\gg k_BT$.
As it is shown below, one can neglect the energy dependence
of the density of states if $d \ll \xi$.
In this case the derivation of the diagonal conductivity
can be done in a fashion similar to the calculations
for the three dimensional case\cite{efros85+}. It gives
\begin{equation}
  \sigma_{xx}(\omega) = \frac{\pi^2}{4}
\frac{e^2}{\hbar}
\hbar\omega
\left(\hbar\omega + U(r_{\omega})
\right)
g^2(0) \xi r_{\omega}^3,
\label{ModMott}
\end{equation}
where $r_{\omega} = \xi \ln (\Delta_{\xi}/\hbar\omega)$ is the arm of the
typical pair contributing to the absorption and $U(r)$ is given by
Eq.~(\ref{dipole}). This formula is a
modification of the expression derived by Mott\cite{mott}. The value
of the conductivity for the case of the interacting electrons
exceeds that for the noninteracting electrons by a factor of
$(\hbar\omega + U(r_{\omega}))/\hbar\omega$.
It happens because the interaction enhances the probability for a pair to
be occupied by only  one electron\cite{efros85+}.
At low enough frequencies
$U(r_{\omega}) = \frac{2e^2d^2}{\varepsilon r_{\omega}^3} \gg \hbar\omega$
and, therefore, the conductivity grows linearly with frequency.
At larger frequencies the interaction becomes irrelevant and the
conductivity
depends quadratically on the frequency. Thus, one can find the width of
the conductivity peaks by equating $\sigma_{xx}(\omega)$ to $e^2/h$
or
\begin{equation}
\hbar\omega \simeq \Delta_{\xi}
\label{c1}
\end{equation}
where $\Delta_{\xi}=(g(0)\xi^2)^{-1}$.
This condition yields
\begin{equation}
 \Delta\nu = \left(\frac{\omega}{\omega_0}\right)^{\eta},
\label{frequency+}
\end{equation}
where
\begin{equation}
 \eta=1/2\gamma=\kappa,\quad
\hbar\omega_0=\frac{A_1}{\alpha}\frac{e^2d}{\varepsilon\xi_0^2}
\end{equation}
$A_1$ being a numerical
coefficient of the order of unity. Now we can establish the
condition for neglecting the energy dependence of the density of
states. The width of the peak was determined from the condition
that the energy $\hbar\omega$ is of the order of
$\hbar\omega_0\simeq
\Delta_{\xi} \simeq \frac{e^2d}{\alpha\varepsilon\xi^2}$. It is larger
than the characteristic scale of the energy dependence of the
density of states, $\alpha \frac{e^2}{\varepsilon d}$ if inequality
$d \ll \alpha\xi$ is valid.
Under this condition the energy of the interaction $U(\xi)$ is also
smaller than $\Delta_{\xi}$.

If $d \gg \alpha\xi$, the width of the peak is given\cite{Polyakov} by
Eq.~(\ref{frequency+})
\begin{equation}
 \eta=1/\gamma=\kappa,\quad
\hbar\omega_0={A_2}\frac{e^2}{\varepsilon\xi_0},
\end{equation}
where $A_2$ is a numerical coefficient.

We see again the crossover in the exponent of the frequency dependence
of peak width at  frequency
$\omega_c\simeq\frac{1}{\hbar}\frac{e^2}{\varepsilon d}$.

Comparing the results of this section with those of
Sec.~\ref{sec:temperature}
we notice that $\hbar\omega/k_B$ always plays the role of the ``effective''
temperature in agreement with dynamic scaling approach.

\section{Experiments with gated samples}
\label{sec:experiment}
In this section we briefly discuss the current
situation with the experiments on gated structures.
The first group of the experiments we would like to mention is the
measurements on GaAs heterostructures.
Naturally, it is most convenient  to study
the modification of the width of the peak with screening  in the
case of spin split peaks which are usually quite narrow. Unfortunately,
the gated structures in this regime have not been studied yet.
All the experiments appeared up to now\cite{Jiang,Dahm,Hughes,Friedland}
deal with low-mobility samples where peaks corresponding to the
opposite spin directions are not split. These works are devoted to
the investigation of the insulator - quantum Hall conductor - insulator
transition in 2D electron gas.
According to Khmelnitskii\cite{Khmelnitskii} and Laughlin\cite{Laughlin}
this transition is related to the evolution of  energy
of the isolated extended states with magnetic field. With increasing magnetic
field the energy of the extended state first drifts down and crosses
the Fermi level, changes the direction of drift,
and moves up crossing the Fermi level for
the second time.
In the vicinity
of both critical points the localization length $\xi$ diverges presumably
according to the scaling law (\ref{localization}) and so two distinct peaks
of the conductivity
appear. Conductivity away from the peaks was shown to be due
to the VRH\cite{Jiang,Dahm,Hughes,Friedland}. Therefore, our
theory of the peak width can be applied for this situation, too.

In Refs.\cite{Jiang,Dahm,Hughes,Friedland} the metallic gate was used
to decrease the electron concentration in order to increase the role
of disorder. Therefore,  we can use
the expression (\ref{nu_t_mott}) with $\kappa=1/2\gamma$ to describe
the width of the peak in the low temperature limit. The value
of the exponent $\gamma$ for the spin degenerate case was conjectured
\cite{Polyakov}
to be given by $\gamma=2\gamma_0$ ($\gamma_0\simeq 2.3$).
It leads to $\kappa=1/4\gamma_0\simeq0.11$ in
experimental situation under discussion.
At a particular value of gate voltage two critical points merge, i.e.
the level of the extended state only touches the Fermi level
at some value of magnetic field $H_c$, not intersecting it.
We conjecture that at this point $\gamma=4\gamma_0$ \cite{excuse}
which leads to the
value of the temperature exponent $\kappa=1/8\gamma_0\simeq 0.05$.
Observation of such small values of the exponent $\kappa$ presumably
requires  measurements at  temperatures lower than those of
Refs.\cite{Jiang,Dahm,Hughes,Friedland}. Therefore, we do not
attempt to compare  their data with our predictions.

The other type of gated structures is the Si MOSFETs where the gate is
typically closer to the plane of the 2D electron gas. However, the
values of the exponent $\kappa$ obtained in
Ref.\cite{dolgopolov91,iorio92,wakabayashi} are not consistent with each
other being distributed in a broad range $0.2-0.9$. Therefore, the
comparison of the theory to these experiments seems to be premature.

\section{Conclusion}
\label{conclusion}

In this paper we studied the effects of the screening of the Coulomb
interaction on the dependence of the conductivity peak width on temperature,
current and frequency. The screening results in substantial broadening of the
peaks and, moreover, in change of the corresponding exponents.
The temperature and frequency exponents $\kappa$ and $\eta$
(see Eqs.~(\ref{nu_t}), (\ref{frequency+})) turn out to be
equal to $1/2\gamma$ rather than $1/\gamma$ for the non-screened Coulomb
interaction ($\gamma$ is the localization length exponent).
We expect that the observation of such a modification of the
scaling dependences will prove that the Coulomb interaction plays a
crucial role for the conductivity in the minima and for
the width of the peaks
in the IQHE.

Another important prediction of this paper is the crossover in
the temperature
dependence of the conductivity in the VRH regime. With decreasing
temperature the crossover between the laws $\ln\sigma\propto T^{-1/3}$
and  $\ln\sigma\propto T^{-1/2}$ is predicted to take place
at $T\simeq 0.013\frac{e^2\xi}{\varepsilon d^2}$. The observation of such a
crossover could become a new tool to study the behavior of the localization
length $\xi(\nu)$.

\acknowledgments
We are grateful to A. J. Dahm, S. Girvin,
S. Koch and especially to D. G. Polyakov
for useful discussions and reading of the manuscript and to
A. J. Dahm, R. J. F. Hughes, M. E. Raikh and R. Reintzsch for
providing us with their results prior to the publication.
This work was supported by the National Science
Foundation under Grant No.~DMR-9020587.
\begin{figure}
\caption{The schematic dependence of the width $\Delta\nu$
of the conductivity peak
on the temperature for  different distances $d$
from the plane of 2D electron gas to the gate.
At $d \ll\xi_0$ all the dependence is described by
Eq.~(\protect\ref{nu_t_mott})
whereas at $d\to\infty$ it is described by Eq.~(\protect\ref{nu_t}).
At intermediate values of $d$ crossover between
these two dependences takes place with increasing temperature.}
\end{figure}
\begin{figure}
\caption{Dimensionless functions $f(x)$ (a) and $F(x)$ (b) from
Eqs.~(\protect\ref{wholerange}) and  (\protect\ref{nu_t1}) are depicted
by solid lines.
Their asymptotes given by Eqs.~(\protect\ref{f}) and
(\protect\ref{g_as}) are shown  by
dashed lines.}
\end{figure}

\begin{references}
\bibitem{excuse} One can argue that in this case the role
of $|\nu-\nu_0|$ in Eq.~(\ref{localization}) is played by the
energy distance between the localized state and the Fermi level.
In this case this distance changes as $(H-H_c)^2$ what determines
the new localization length exponent. (This argument belongs to J. T. Chalker).
\end{references}
\end{document}